\documentclass[aps,prl,reprint,superscriptaddress,nofootinbib,floatfix]{revtex4-2}

\usepackage{amsmath,amssymb}
\usepackage{graphicx}
\usepackage{bm}
\usepackage[colorlinks=true,linkcolor=blue,citecolor=blue,urlcolor=blue]{hyperref}

\newcommand{\rd}{r_{\rm d}}
\newcommand{\DM}{D_{\rm M}}
\newcommand{\Dhub}{D_{\rm H}}
\newcommand{\thbao}{\theta_{\rm BAO}}
\newcommand{\zeff}{z_{\rm eff}}
\newcommand{\MM}{{\rm MM}}
\newcommand{\Ntw}{{\rm N20}}

\begin{document}

\title{A blind spot in transverse BAO calibration}

\author{Domenico Sapone}
\email{domenico.sapone@uchile.cl}
\affiliation{Departamento de F\'isica, FCFM, Universidad de Chile, Santiago, Chile}

\date{\today}

\begin{abstract}
Transverse baryon acoustic oscillation (BAO) measurements are increasingly used
for cosmological inference, and carry a calibration that no such inference can
constrain. A constant error in the transverse BAO scale is exactly degenerate with
the combination $\rd h$: it leaves the goodness of fit unchanged and the recovered
parameters plausible, and is therefore invisible to any analysis that uses these
measurements alone. The radial BAO sector removes this degeneracy, supplying
$\DM/\rd$ by integration without reference to $H_0$ or to any model for the
expansion rate. In flat
Friedmann--Lema\^itre--Robertson--Walker (FLRW) geometry the relation between the
two sectors is an identity, so the sound horizon and the dark-energy equation of
state cancel as well. An integrated form of this identity reduces the consistency
test to a straight line, whose slope measures a relative transverse calibration
$\varepsilon$. A departure from $\varepsilon = 1$ cannot be produced by any
dark-energy model, nor by spatial curvature: it indicates an
inconsistency in the measurement chain rather than in the cosmology. We apply the
test to the two main SDSS transverse BAO compilations, which give
$\varepsilon = 1.073 \pm 0.021$ and $1.021 \pm 0.029$. The first differs from
unity at $3.8\sigma$ using the published independent errors, while the second is
consistent with unity. The two compilations themselves differ by
$(5.4 \pm 1.4)\%$, or $3.9\sigma$, and the offset is constant in redshift. The test provides a direct diagnostic for current and
future angular BAO measurements.
\end{abstract}

\maketitle

\textit{Introduction.}---Baryon acoustic oscillations (BAO) provide a standard
ruler of comoving length $\rd$, the sound horizon at the baryon drag epoch.
Measured transverse to the line of sight, the acoustic scale subtends
$\thbao(z) = \rd/\DM(z)$, while in the radial direction it spans
$\Delta z = \rd/\Dhub(z)$, with $\Dhub(z) \equiv c/H(z)$. The two measurements
therefore probe the same ruler through two different distance functions.

The standard three-dimensional (3D) BAO analysis converts angular and redshift
coordinates into comoving coordinates using a fiducial cosmology and measures
radial and transverse dilation factors \cite{Weinberg:2013agg,SDSS:2005xqv,%
2dFGRS:2005yhx,Blake:2011en,BOSS:2012dmf,eBOSS:2015jyv,DESI:2025zgx}. An
alternative approach measures the BAO feature directly in the angular correlation
function $w(\theta)$ in thin redshift shells \cite{Sanchez:2010zg,%
Carvalho:2015ica,Menote:2021jaq}. This transverse-only approach makes fewer
assumptions about the background cosmology, but its uncertainties are larger.

Recent analyses have reported a $3.5$--$3.7\sigma$ difference between transverse
BAO measurements and the transverse distances inferred from DESI
\cite{Favale:2024sdq,Pantos:2026rpe}, while other analyses find consistency
\cite{Sabogal:2025qhz,Menote:2021jaq}. These apparently different conclusions use
different transverse BAO catalogues.

These comparisons all use the transverse sector of the 3D measurements. It was shown in~\cite{Pantos:2026rpe} that the published 3D distances are
fiducial-independent by construction due to the product $\alpha_\perp \DM^{\rm fid}/\rd^{\rm fid}$, with residual at $\lesssim 0.3\%$. Comparing the transverse
scales with the transverse sector of the 3D measurements, they further showed that
no CPL model reconciles the two datasets, and that no smooth modification of
$\DM(z)$ removes the disagreement at $z = 0.510$. The alternative is not a modification of $\DM(z)$ at all: a constant recalibration of $\thbao$
leaves the geometry untouched and shifts every transverse distance by the same
factor.

The radial BAO sector provides the reference that breaks this degeneracy without
requiring a model for the expansion history, using the same geometric relation
that underlies the null tests developed to probe FLRW geometry and cosmological
models
\cite{Clarkson:2007pz,Garcia-Bellido:2008xmz,Clarkson:2012bg,Seikel:2012cs,%
Nesseris:2010ep,Heavens:2011mr,Sapone:2014nna,Nesseris:2014mfa,%
Nesseris:2014qca,Laurent:2016eqo,Chiang:2017yrq,vonMarttens:2018iav,%
vonMarttens:2020apn,Euclid:2021frk,Andrade:2021njl,Millard:2026wnd,%
Sapone:2026iwc,Martinelli:2026wjp}. Here we use the same geometric relation for a
different purpose: as a data-level consistency and calibration test.

Applying the same relation to the anisotropic sectors of DESI DR2, it was found in~\cite{Sapone:2026iwc} that $\mathcal{C}_0 = -0.01 \pm 0.01$: the transverse and radial 3D measurements
are mutually calibrated within present precision. The
departure we find here is therefore located in the transverse-only measurements
rather than in the radial reference.

\textit{BAO consistency function.}---In flat FLRW geometry the transverse comoving
distance satisfies
\begin{equation}
    \DM'(z) = \Dhub(z),
    \label{eq:DMprime}
\end{equation}
for any expansion history. Defining $d(z) \equiv \DM(z)/\rd$ and
$d_H(z) \equiv \Dhub(z)/\rd$, Eq.~\eqref{eq:DMprime} gives
\begin{equation}
    \mathcal{N}(z) \equiv \frac{d'(z)}{d_H(z)} - 1 = 0 .
    \label{eq:null}
\end{equation}
This is the flat limit of the Clarkson--Bassett--Lu consistency relation
\cite{Clarkson:2007pz}. Neither $w(z)$ nor $H_0$ appears in Eq.~\eqref{eq:null},
and $\rd$ cancels.

Direct numerical derivatives of sparse BAO data are noisy, so we use the
integrated form of Eq.~\eqref{eq:null}. We allow the measured transverse angular
scale to differ from the ideal relation by a constant factor,
$\thbao(z) = \varepsilon\,\rd/\DM(z)$, and define
\begin{equation}
    T(z) \equiv \frac{180/\pi}{\thbao(z)}\,,
    \qquad
    X(z) \equiv \int_{z_0}^{z}\frac{\Dhub(\tilde z)}{\rd}\,d\tilde z\,,
    \label{eq:TX}
\end{equation}
where $\thbao$ is expressed in degrees. Equation~\eqref{eq:null} then becomes
\begin{equation}
    T(z) = A + B\,X(z)\,,
    \label{eq:master}
\end{equation}
with $B = 1/\varepsilon$ and $A = \DM(z_0)/(\rd\,\varepsilon)$.
Equation~\eqref{eq:master} is the main result of this work. The transverse data
enter only through $T$ and the radial data only through $X$, so the relation is a
straight line whose slope measures the relative calibration of the two sectors. If
both $A$ and $B$ are free, the null hypothesis is simply that $T$ is linear in
$X$; this is an anchor-free test and requires no external distance normalization.
If $\DM(z_0)/\rd$ is fixed independently, the slope gives $\varepsilon$, which becomes a measurement. 

The construction is independent of the dark-energy model. A change in $w(z)$
changes $\DM$ and $\Dhub$ together and preserves Eq.~\eqref{eq:DMprime}, and a
constant rescaling of $\rd$ cancels from the relation. A constant departure in
$\varepsilon$ is therefore not degenerate with the expansion history in this test.
It measures instead a relative calibration or standard-ruler inconsistency between
the transverse and radial measurements. 

Possible sources of such an offset include projection corrections associated with
finite shell width \cite{Sanchez:2010zg,Crocce:2010qi}, the transverse part of the
nonlinear BAO shift \cite{Eisenstein:2006nj,Seo:2009fp}, and biases associated
with the template used to locate the BAO peak \cite{Menote:2021jaq}. The nonlinear
anisotropic shift is a few $\times 10^{-3}$ for the scales considered here, well
below the offsets found below. Only the template bias is known to be exactly
constant in redshift; the others vary weakly across the narrow range spanned by
each compilation, and we fit each compilation separately for this reason.

The test also responds to spatial curvature, to departures from FLRW geometry, to
redshift evolution of the standard ruler, and to systematics that affect the two
sectors differently. For spatial curvature,
$(d'/d_H)^2 = 1 + K d^2$ with $K = \Omega_k (H_0\rd/c)^2$, so the effect grows
with redshift. A constant $\varepsilon$ is therefore distinct from the curvature
contribution.

Finally, the assumptions underlying the BAO standard ruler have been enumerated
explicitly, and the first of them is isotropy, dismissed as conservative since a
ruler set by early-time physics could not violate it without violating the
Cosmological Principle \cite{Brieden:2022heh}. That argument concerns the ruler
itself. It does not cover the case in which the ruler is isotropic but the
transverse and radial \emph{measurements} of it are not, which is what
$\varepsilon$ measures.

\textit{Data.}---We use the radial BAO measurements from the BOSS DR12 consensus
\cite{BOSS:2016wmc} at $z = 0.38, 0.51, 0.61$, together with DESI DR2
\cite{DESI:2025zgx} above $z = 0.61$. 

The normalization for the anchored fit is the BOSS transverse distance
$\DM(0.38)/\rd = 10.231 \pm 0.166$, taken at $z_0 = 0.38$, the lowest redshift
with an anisotropic BAO measurement~\cite{BOSS:2016wmc}. Its uncertainty is
propagated jointly with the radial covariance, since it comes from the same
analysis as two of the radial nodes. To evaluate $X(z)$ we interpolate $\ln(\Dhub/\rd)$ as
a cubic function of $\ln(1+z)$, repeat the calculation with quadratic and quartic
interpolants, and include the spread as an interpolation systematic.

Throughout, transverse measurements are quoted in the units in which they were
published: $\thbao$ in degrees for the angular analyses, $\DM/\rd$ for DES~Y6 and
for the three-dimensional measurements. Equation~\eqref{eq:TX} converts between
them.

The transverse sector contains two SDSS-based compilations. Menote \& Marra (MM)
\cite{Menote:2021jaq} provide 14 measurements over $0.35 \le z \le 0.63$ from BOSS
DR12 and eBOSS DR16, together with a published $14\times14$ correlation matrix.
Their analysis includes a correction for a template inference bias measured from
1000 mocks, $B_{\rm inf} = (3.2 \pm 0.1)\%$. Using their public data we verify
that the ratio of the pre- and post-correction angular scales equals $1.032$ at
all fourteen redshifts, so the correction is applied uniformly. The Nunes
\textit{et al.} (N20) compilation \cite{Nunes:2020hzy} contains 12 measurements
over $0.365 \le z \le 0.65$, based on SDSS DR10 and DR11
\cite{Carvalho:2015ica,Alcaniz:2016ryy,Carvalho:2017tuu}, and does not apply an
equivalent correction.

The MM values and covariance are adopted unchanged in the more recent
compilation~\cite{Sabogal:2025qhz}, which adds measurements at $z = 0.11$
\cite{deCarvalho:2021azj}, $z = 1.725$ \cite{Avila:2025qxc} and $z = 2.225$
\cite{deCarvalho:2017xye} and contains none of the older measurements entering
N20. We use a similar set outside the SDSS footprint, omitting $z = 0.11$, which
lies below the lowest radial measurement, and adding the second quasar shell at
$z = 1.775$ and DES~Y6 at $\zeff = 0.85$, $\DM/\rd = 19.51 \pm 0.41$
\cite{DES:2024pwq}, from a southern footprint independent of SDSS imaging. For the
quasar shells we adopt the revised values, $\thbao = 1.911 \pm 0.062$ and
$1.727 \pm 0.081$ degrees.
Adding these points to either SDSS compilation would tie the two fits to a common
set of measurements and hide the difference between the catalogues, so we treat
them as a separate block.

\begin{figure*}[t]
\centering
\includegraphics[width=\textwidth]{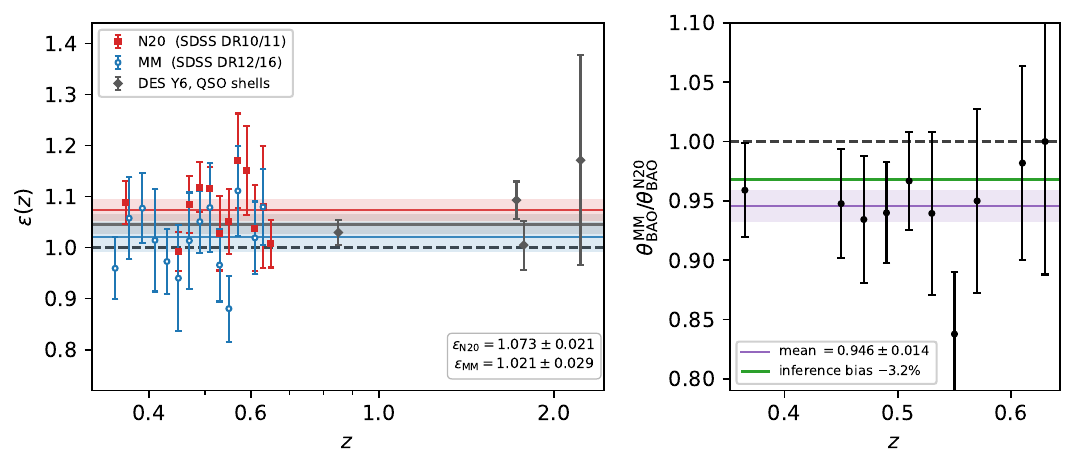}
\caption{Left panel: the transverse calibration
$\varepsilon(z) = [\DM(z)/\rd]_{\rm radial}\,\thbao(z)$ for the N20 (red squares) and MM (blue circles) compilations; grey diamonds are the non-SDSS measurements. Bands show the fitted constant $\varepsilon$ for each block. The MM uncertainties are inflated by the factor $3.3$ discussed in the text. Right panel: the two compilations compared directly at the ten redshifts where both provide a measurement. The purple band is the weighted mean, the green line the $3.2\%$ template inference bias, corrected in MM and not in N20.}
\label{fig:main}
\end{figure*}

\textit{Results.}---We first perform the anchor-free test, fitting
Eq.~\eqref{eq:master} with both $A$ and $B$ free. The N20 data give
$\chi^2/{\rm dof} = 11.6/10$ ($p = 0.31$) and MM gives $11.2/12$ ($p = 0.52$).
Neither compilation shows evidence for a departure from the redshift dependence
predicted by the flat-FLRW relation. In particular, their redshift dependence does
not require a change in the expansion history. With $A$ free the intercept and
slope are strongly anti-correlated, so this test constrains linearity rather than
$\varepsilon$ itself.

We next use the anchor to determine the relative calibration:
\begin{align}
    \varepsilon^{\Ntw} &= 1.073 \pm 0.021 \,, \label{eq:epsN20}\\
    \varepsilon^{\MM}  &= 1.021 \pm 0.029 \,. \label{eq:epsMM}
\end{align}
Using the published N20 errors the first differs from unity by $3.8\sigma$, while
MM is consistent with unity at $0.7\sigma$. The MM uncertainty in
Eq.~\eqref{eq:epsMM} includes an inflation factor justified below; without it the
central value is almost unchanged, $\varepsilon^{\MM} = 1.024 \pm 0.015$. The
central values are also stable against the interpolation of the radial data:
quadratic and quartic interpolants change $\varepsilon$ by less than the quoted
statistical uncertainty. The left panel of Fig.~\ref{fig:main} shows the two
determinations point by point.

The main limitation is the covariance between the N20 measurements. The twelve
points use the same SDSS imaging footprint, but no covariance for a possible
common angular systematic is published. We therefore repeat the fit after adding a
common correlation $\rho\sigma_i\sigma_j$ between them. The fitted central value
changes by less than $0.001$ over $0 \le \rho \le 0.5$, while the significance
changes from $3.8\sigma$ for independent points to $2.7\sigma$ for $\rho = 0.3$.
We quote the result using the published errors, stressing that the central
calibration offset is more robust than its precise significance.

Finally, the N20 compilation combines three separate analyses. Fitting each
independently gives $\varepsilon = 1.089 \pm 0.042$, $1.071 \pm 0.024$ and
$1.070 \pm 0.035$, and removing any one of them leaves $\varepsilon$ within
$1.071$--$1.078$. The offset is therefore a property of the analysis approach and
not of a single measurement.

The two transverse catalogues can also be compared without the radial reference.
At the ten redshifts where both provide a measurement, right panel of
Fig.~\ref{fig:main},
\begin{equation}
    1 - \thbao^{\MM}/\thbao^{\Ntw} = (5.4 \pm 1.4)\% \,,
    \label{eq:cataloguediff}
\end{equation}
so the two catalogues differ from one another at $3.9\sigma$, using the observed
scatter of the ratios as the uncertainty. This result does not depend on the
external anchor. The ratio of the two fitted calibration factors,
$\varepsilon^{\Ntw}/\varepsilon^{\MM} = 1.051$, is consistent with it. The two
determinations are not independent, since they share the anchor and the radial
reference, and their difference is $0.052 \pm 0.030$ once this is propagated; the
direct comparison in Eq.~\eqref{eq:cataloguediff} is the sharper of the two
because its uncertainty comes from the observed scatter of the ten ratios and does
not involve the quoted errors of either compilation. In terms of the calibration,
$\varepsilon^{\Ntw} - 1 = 7.3\%$ decomposes into $5.2\%$ from the catalogue
difference and $2.1\%$ common to both compilations: with the MM calibration, N20
would give $\varepsilon = 1.02$, consistent with unity.

\textit{An independent check.}---The MM compilation provides an independent
indication that its published errors do not describe the observed scatter. We fit
the MM measurements directly with flat $\Lambda$CDM, using only $\Omega_m$ and
$\rd h$ as free parameters and no radial BAO information. With their published
covariance we obtain
\begin{equation}
    \chi^2/{\rm dof} = 127.5/12 \,, \qquad p = 1.9\times10^{-21}\, .
    \label{eq:MMchi2}
\end{equation}
The dominant residuals occur at $z = 0.55$ and $z = 0.57$, where the reported
angular scales differ by about $22\%$ despite individual uncertainties of about
$2\%$. Removing the most discrepant point still leaves $p \sim 10^{-11}$, and
using only the diagonal covariance gives $\chi^2 = 118.3$, so the off-diagonal
terms are not responsible.

Using the public MM data files we reproduce their likelihood and published
constraint $\rd h = 97.5 \pm 3.6$~Mpc to within $0.03\sigma$. The diagonal of the
covariance agrees with the published $\sigma_{\rm BAO}^2$ and the likelihood
contains no terms beyond the Gaussian quadratic form, so the large $\chi^2$ is not
caused by a different likelihood implementation. Furthermore, fitting the same two-parameter
model to the seventeen-point compilation in~\cite{Sabogal:2025qhz}, which
adopts the MM values and covariance together with three additional measurements,
gives $\chi^2/{\rm dof} = 129.2/15$ ($p = 3\times10^{-20}$), with residuals of
$-6.6\sigma$ at $z = 0.55$ and $+4.2\sigma$ at $z = 0.39$. Our best fit,
$\Omega_m = 0.46$ and $\rd h = 96.5$~Mpc, is consistent with their published
$\Omega_m = 0.421^{+0.073}_{-0.10}$ and $\rd h = 98.1^{+2.9}_{-2.6}$~Mpc, so this
is not a difference of implementation. The MM residuals are large but change sign, so they largely cancel in the mean: the compilation is unbiased relative to the radial sector but imprecise. Its inflated uncertainty admits $\varepsilon$ between $0.96$ and $1.08$ at $2\sigma$, which includes $\varepsilon^{\Ntw}$. The sharper statement is therefore the direct catalogue comparison of Eq.~\eqref{eq:cataloguediff}, not the difference between the two fitted values of $\varepsilon$.

The same excess appears in our null test: with the published MM errors the
anchored fit gives $\chi^2/{\rm dof} = 122.3/13$. The two tests identify the same
scatter by different routes, one comparing the transverse measurements with a
cosmological distance curve and the other with the radial BAO relation, without
assuming an expansion history. Following the standard treatment of underestimated
errors \cite{ParticleDataGroup:2024cfk}, the MM uncertainties would need to be
inflated by $\sqrt{127.5/12} \simeq 3.3$ to describe the observed scatter, and we
apply this inflation in Eq.~\eqref{eq:epsMM}. The N20 compilation passes the same
test, $\chi^2/{\rm dof} = 11.8/10$ ($p = 0.30$), and there is no need to inflate them.

\begin{figure}[t]
\centering
\includegraphics[width=\columnwidth]{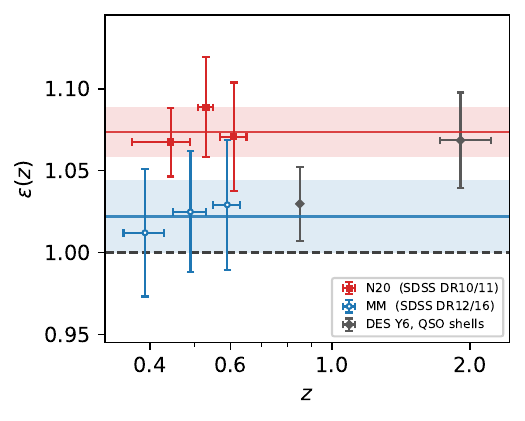}
\caption{$\varepsilon$ fitted independently in redshift bins. Horizontal bars span
each bin and the bands show the weighted mean for each SDSS compilation. The
offset is constant in redshift for both, as expected for a calibration effect. The
anchor variance is excluded from the per-bin errors.}
\label{fig:binned}
\end{figure}

\textit{Redshift dependence.}---A constant calibration is of particular interest
because it is not expected from a change in the expansion history or from spatial
curvature. We therefore fit $\varepsilon$ independently in redshift bins, shown in Fig.~\ref{fig:binned}. Against a constant, N20 gives $\chi^2/{\rm dof} = 0.34/2$
($p = 0.84$) and MM gives $0.10/2$ ($p = 0.95$). Including the non-SDSS points
extends the range to $z \simeq 1.9$ and gives $p = 0.55$ and $p = 0.77$. The
fitted linear drifts are $d\varepsilon/dz = -0.005 \pm 0.023$ for N20 and
$+0.034 \pm 0.025$ for MM, and an unbinned linear-drift fit changes $\chi^2$ by
$0.00$ and $1.74$ respectively for one extra parameter. The external anchor is the
same in every bin, so its uncertainty shifts all bins together and cannot create a
false evolution; we remove it from the per-bin errors.

The absence of significant redshift evolution is consistent with a calibration
effect, and distinguishes the result from the leading geometric effect of spatial
curvature, which grows as $d(z)^2$. Fitting $K$ jointly with $\varepsilon$ gives
$\Omega_k$ consistent with flatness for both compilations, with
$\Delta\chi^2 < 2$ between the flat and best-fitting curved cases. The data
therefore do not show evidence that the constant offset is produced by the
expansion history or by spatial curvature.

There is a possible connection with the template treatment in the two
compilations. MM corrects a template inference bias of $(3.2 \pm 0.1)\%$, while
the N20 analyses do not apply an equivalent correction. This has the same sign and
a comparable size to part of the observed $5.4\%$ catalogue difference. However,
the bias was measured for the MM template, shell width and sample, and the N20
pipeline differs in all three. We therefore do not assume that the correction
transfers to N20, and do not claim that it explains the full difference. It is
instead an example of a pipeline effect that can produce a constant transverse
offset. The non-SDSS measurements give
$\varepsilon^{\rm non\text{-}SDSS} = 1.046 \pm 0.020$, between the two SDSS
determinations. These measurements use different templates, so the MM correction does not apply to them; if a template bias of this kind were the only systematic affecting angular BAO, they should still return $\varepsilon= 1$. That they sit $2.5\sigma$ above unity suggests a residual offset, though four measurements cannot establish whether it is common to the method or coincidental.


\textit{The blind spot.}---The importance of the test can be seen directly from a
transverse-only cosmological fit. In flat $\Lambda$CDM,
\begin{equation}
    \thbao(z) \propto
    \frac{\rd h}{\displaystyle\int_0^z dz'/E(z')} ,
    \label{eq:degeneracy}
\end{equation}
so a constant rescaling of $\thbao$ has exactly the same effect as a rescaling of
$\rd h$. The degeneracy between $\rd h$ and the overall distance scale is of
course well known; what it implies for calibration is less often noted. A
calibration error in $\thbao$ enters the fit in exactly the same way as a shift in
$\rd h$, so it produces neither a poor fit nor an implausible parameter value, and
leaves no signature by which it could be identified.

The consequence is quantitative, and we verified it directly on the data used
here. Fitting $\Omega_m$ and $\rd h$ to the two compilations separately gives
$\rd h = 97.7^{+3.5}_{-3.8}$~Mpc for MM and $104.8^{+7.6}_{-8.2}$~Mpc for N20,
both compatible with Planck at below $1\sigma$ in both parameters, despite the
difference between them. The correlation coefficient between
$\Omega_m$ and $\rd h$ is $\simeq -0.97$ in both fits. Dividing the N20 angular scales by the fitted $\varepsilon^{\Ntw}$ and refitting leaves $\chi^2/{\rm dof} = 11.9/11$ unchanged, while $\rd h$ moves from $104.8$ to $97.8$~Mpc, i.e.\ by exactly that factor. Combining transverse BAO with an external probe does not
resolve this. The normalisation is then set by the external dataset, and a nonzero
$\varepsilon$ biases it while leaving its uncertainty unchanged. What survives
uncontaminated is the shape of $\DM(z)$, and hence $\Omega_m$, provided
$\varepsilon$ is constant.

A uniform rescaling of the BAO measurements has been considered as a hypothetical
systematic and shown to shift $H_0$ while leaving $\Omega_m$ unchanged
\cite{Pedrotti:2025ccw}. Being isotropic, it cancels identically in the comparison
constructed here; the same work notes that fitting the rescaling as a free
parameter would leave the measurements as unanchored probes of the shape of the
expansion rate, and fixes it for that reason. Equation~\eqref{eq:master} removes
this limitation, because $\DM/\rd$ is not computed from a model but reconstructed
by integrating the measured radial sector, so $h$ never enters. This is the sense
in which the radial data supply information that is genuinely independent of the
combination $\rd h$.


\textit{Conclusions.}---A constant calibration error in the transverse BAO scale
cannot be detected by a cosmological fit to those measurements alone. We have
constructed a model-independent null test that closes this gap by relating
transverse and radial BAO measurements. In flat FLRW geometry the relation is independent of the sound
horizon, $H_0$ and the dark-energy equation of state, and its integrated form
reduces the comparison to a straight line whose slope measures a relative
transverse calibration.

The two main SDSS transverse compilations do not give the same calibration when
compared with the same radial reference: N20 is offset by about $7\%$
($3.8\sigma$), while MM is consistent with unity. Compared directly
with each other the two catalogues differ at $3.9\sigma$ and the offset is constant in redshift. An independent test of the MM catalogue gives
$\chi^2/{\rm dof} = 127.5/12$, showing that its published uncertainties do not
describe its own point-to-point scatter; the same excess is found by the null
test. The same excess is present in the seventeen-point compilation currently in
use \cite{Sabogal:2025qhz}, $\chi^2/{\rm dof} = 129.2/15$, since it adopts the MM
values and covariance unchanged. Removing the older measurements resolved the
calibration offset, which is why MM is consistent with the radial sector, but it
did not address the error budget: a parameter-level analysis cannot see either
problem. This difference helps explain why analyses using N20 report a
$3.5$--$3.7\sigma$ tension while analyses using MM find consistency. The two
conclusions cannot be interpreted only as different cosmological models; they also
depend on the transverse catalogue.

The result should not be read as a detection of new physics. The test is
insensitive to $w(z)$, and a constant offset is not the expected signature of
spatial curvature. The appropriate interpretation is a relative calibration or
standard-ruler inconsistency between the two BAO sectors. 

The next step is a uniform reanalysis of the underlying SDSS transverse data with
a common pipeline, a common mock-based calibration, and an error model validated
against the observed scatter. This matters because such an error propagates
directly into dark-energy and Hubble-constant constraints. More generally, Eq.~\eqref{eq:master} can be applied
whenever transverse and radial BAO measurements overlap in redshift, and because
it requires no model for $H(z)$, it can be used as a data-level consistency check
before parameter inference. Applied here, the same relation identified two
distinct problems: a constant calibration offset between one compilation and the
radial sector, and an excess scatter within the other that its published errors do
not describe. A consistency relation of this kind is therefore not only a test of
cosmological models, but a diagnostic that can be applied to the data themselves,
before any parameter inference.

\begin{acknowledgments}
DS acknowledges financial support from Fondecyt Regular N.~1251339.
\end{acknowledgments}

\bibliography{references}

\end{document}